\documentclass[3p,twocolumn]{elsarticle}
\biboptions{sort&compress}

\usepackage{amsmath}
\usepackage{amsfonts}
\usepackage{color}
\usepackage{float}
\usepackage{graphicx}
\usepackage[caption = false]{subfig}
\usepackage{chemformula}

\makeatletter
\def\ps@pprintTitle{%
  \let\@oddhead\@empty
  \let\@evenhead\@empty
  \def\@oddfoot{\reset@font\hfil\thepage\hfil}
  \let\@evenfoot\@oddfoot
}
\makeatother

\begin{document}
\title{Position control of an acoustic cavitation bubble by reinforcement learning}%

\author[hdr]{K\'alm\'an Klapcsik\corref{cor1}}
\ead{kklapcsik@hds.bme.hu}

\author[vik]{B\'alint Gyires-T\'oth}
\ead{toth.b@tmit.bme.hu}

\author[ljub]{Juan Manuel Rossell\'o}
\ead{jrossello.research@gmail.com}

\author[hdr]{Ferenc Heged\H{u}s}%
\ead{fhegedus@hds.bme.hu}

\cortext[cor1]{Corresponding author}
\address[hdr]{Department of Hydrodynamic Systems, Faculty of Mechanical Engineering, Budapest University of Technology and Economics, M\H{u}egyetem rkp. 3., H-1111 Budapest, Hungary}
\address[vik]{Department of Telecommunications and Media Informatics, Faculty of Electrical Engineering and Informatics, Budapest University of Technology and Economics, M\H{u}egyetem rkp. 3., H-1111 Budapest, Hungary}
\address[ljub]{Faculty of Mechanical Engineering, University of Ljubljana, A\v{s}ker\v{c}eva 6, 1000 Ljubljana, Slovenia}

\date{\today}%

\begin{abstract}
A control technique is developed via Reinforcement Learning that allows arbitrary controlling of the position of an acoustic cavitation bubble in a dual-frequency standing acoustic wave field. The agent must choose the optimal pressure amplitude values to manipulate the bubble position in the range of $x/\lambda_0\in[0.05, 0.25]$. To train the agent an actor-critic off-policy algorithm (Deep Deterministic Policy Gradient) was used that supports continuous action space, which allows setting the pressure amplitude values continuously within $0$ and $1\, \mathrm{bar}$. A shaped reward function is formulated that minimizes the distance between the bubble and the target position and implicitly encourages the agent to perform the position control within the shortest amount of time. In some cases, the optimal control can be 7 times faster than the solution expected from the linear theory. 
\end{abstract}


\maketitle


\section{Introduction}

The irradiation of the liquid domain with high-frequency and high-intensity ultrasound results in the forming of thousands of micron-sized radially pulsating bubbles. The collapse of these bubbles induces chemical reactions, which is the keen interest of sonochemistry \cite{Yasui2018}. Despite the great potential of various applications \cite{Xu2013, Sivasankar2009a, Sivasankar2009b, Gole2018, Pradhan2010, Gedanken2004}, the biggest challenge, the scale-up of applications feasible for industrial sizes, is still unsolved \cite{Sutkar2009}. The attenuation of the sound waves in the densely packed bubble clusters is one of the main limitations \cite{vanIersel2008, Sojahrood2017} and the spurious interaction of the bubbles with the container. Controlling the dynamics of bubble clusters via acoustic manipulation methods; for example, the positioning of bubbles within the clusters, can be a possible solution to overcome such limitations.

Although acoustic manipulation methods are extensively used for solid particles in various applications, such as particle manipulation \cite{Sriphutkiat2017, DRON20131280, Kandemir2021, Kourosh2020}, pattern formation \cite{Vuillermet2016} and micro-assembly \cite{Goldowsky2013}, the utilization of such control techniques is not widespread in the literature in the case of bubble clusters. Only elementary techniques have been applied to cluster control, such as on/off control \cite{Maeda2021, Lee2011} or bi-frequency driving to avoid spatial instability \cite{Rosello2016, Rosello2015}.

The traditional acoustic manipulation devices create a simple pattern of standing waves in a chamber or channel \cite{Sriphutkiat2017}. In such patterns, fixed trapping points (or lines) exist, e.g., pressure nodes and antinodes, where the mean acoustic radiation force is zero. The external forces drive the particles to these fixed positions. Either by controlling the phase or intensity of the transducers or by switching between resonance modes, the wave field is transformed, and the trapping points can be modified; thus, the position of the particles can be controlled \cite{DRON20131280, Kandemir2021}. A possible method of particle manipulation is the application of phase-controllable ultrasonic standing waves. In this case,  a pair of ultrasonic transducers is applied, and the relative phase between the generated sinusoidal signal is varied to change the position of nodes and antinodes. Abe et al. \cite{Abe2002} achieved the control of bubble motion by using this technique for a single bubble in an acoustic standing wave. 

Simple control methods are possible for bubbles; however, these techniques are limited to weak pressure amplitude, when the bubbles are trapped by either the node or the antinode. The application of high-intensity, multi-frequency driving allows a more general solution; however, the translational motion becomes much more complex; e.g., the bubble may exhibit periodic and chaotic translational oscillations \cite{Mei1991, Feng1995, Doinikov2004a, Mettin2009}, or the bubble can break up \cite{Versluis2010}. Therefore, position control may require more complex manipulation of the acoustic field \cite{Lee2011, Bai2014, Rosello2015, Maeda2021}, which might be well beyond simple intuitions.

The aim is to seek robust control that allows arbitrarily positioning a bubble in a standing wave field, by using reinforcement learning (RL). Reinforcement learning can be seen as a method to control nonlinear systems, which does not rely on analytical knowledge of the underlying dynamical systems. A large number of RL algorithms have been developed in the last decade. These advanced RL algorithms use (deep) neural networks as function approximators that allow them to operate over discrete and continuous state and action spaces; therefore, complex control tasks can be solved \cite{sutton2018reinforcement, jeanfrancois2003markov, Mnih2013, Mnih2015, Lillicrap2016}. In the present paper, this kind of manipulation is achieved by applying dual-frequency excitation \cite{Tatake2002, Suo2018, Zhang2016, Zhang2017, Zhang2015b} and tuning the pressure amplitude values in discrete timesteps.

\section{Mathematical Model}
The coupled radial and translational motion of an acoustic cavitation bubble is described by two coupled ordinary differential equations \cite{Doinikov2002}. The first one is the Keller--Miksis equation \cite{Keller1980}, which is a second-order nonlinear differential equation that describes the radial oscillation of a spherical bubble. The equation is written as

\begin{multline} \label{Eq:KellerMiksis}
	\left( 1-\frac{\dot{R}}{c_{L}} \right) R\ddot{R} +  \left( 1-\frac{\dot{R}}{3c_{L}} \right) \frac{3}{2}\dot{R}^{2} \  = \\
	\left( 1 +  \frac{\dot{R}}{c_{L}} +\frac{R}{c_{L}}\frac{d}{dt} \right)\frac{\left( p_{L} -p(x,t) \right)}{\rho_{L}}+\frac{\dot{x}^2}{4},
\end{multline}

\noindent where $R$ and $x$ are the instantaneous bubble radius and its position, respectively. Furthermore, $\rho_{L}=998\, \mathrm{kg/m^3}$ is the liquid density, $c_L=1500\, \mathrm{m/s}$ is the speed of sound and $p(x,t)$ is the pressure at the centre of the bubble. The dots stand for the derivative with respect to the time. The liquid pressure at the bubble wall is given as

\begin{equation} \label{Eq:BubbleInterior}
	p_{L}= p_G - \frac{2 \sigma}{R} - 4 \mu_{L} \frac{\dot{R}}{R},
\end{equation}

\noindent where $\sigma=0.0725\, \mathrm{N/m}$ and $\mu_{L}=0.001\, \mathrm{Pa\,s}$ are the surface tension and the liquid dynamic viscosity, respectively. The gas pressure $p_G$ is assumed to obey a polytropic state of change

\begin{equation}
	p_G = \left(\frac{2 \sigma}{R_{0}}  + P_{\infty} \right) \left( \frac{R_{0}}{R}  \right)^{3n},
\end{equation}

\noindent where $n=1.4$ is the polytropic exponent and $R_0=60\, \mathrm{\mu m}$ is the equilibrium bubble radius.

The translational motion of the bubble is described as \cite{Doinikov2002, Doinikov2005, Mettin2009}

\begin{equation} \label{Eq:TranslationalMotion}
	R \ddot{x}+3\dot{R}\dot{x}=\frac{3F_{ex}(x,t)}{2\pi\rho_{L}R^2},
\end{equation}

\noindent where $F_{ex}(x,t)$ is the sum of the instantaneous external forces acting on the bubble; namely, the primary Bjerknes force \cite{Crum1975}

\begin{equation}
	F_{B1} = -V(t)\nabla p(x,t),
\end{equation} 

\noindent and the drag force \cite{Levich1962}

\begin{equation} \label{Eq:DragForceLevich}
	F_{D} = -12 \pi \mu_{L} R \left( \dot{x} -v_{ex}(x, t) \right).
\end{equation}

\noindent In the above equations, $V(t)$ is the volume of the bubble, $\nabla p(x,t)$ is the pressure gradient and $v_{ex}(x,t)$ is the velocity induced by the acoustic radiation. From Eqs.~(\ref{Eq:TranslationalMotion})-(\ref{Eq:DragForceLevich}), one can observe that the manipulation of the acoustic field properties allows the position control of the bubble. 

The acoustic field is assumed to be the sum of two standing waves

\begin{equation} \label{Eq:PressureField}
	\begin{split}
		p(x,t) = P_{0} &+ P_{A0}\sin(k_0x)\sin(\omega_0t) \\
		&+P_{A1}\sin(k_1x)\sin(\omega_1t),
	\end{split}	
\end{equation}

\noindent where $P_{0}=1\, \mathrm{bar}$ is the ambient pressure, $P_{A0}$ and $P_{A1}$ are the pressure amplitudes, $\omega_0=2\pi f_0$ and $\omega_1=2\pi f_1$ are the angular frequencies, $k_0=2\pi/\lambda_0$ and $k_1=2\pi/\lambda_1$ are the wavenumbers. Note that $\lambda_0$ and $\lambda_1$ are the wavelengths correspond to the excitation frequencies $f_0=25\, \mathrm{kHz}$ and $f_1=50\, \mathrm{kHz}$, respectively.

The mathematical model is solved numerically by introducing dimensionless variables; namely, the dimensionless bubble radius $y_1=R/R_0$, the dimensionless position $\xi=x/\lambda_0$ and their derivatives with respect to the dimensionless time $\tau=t/(2\pi/\omega_0)$. The dimensionless system of equations is given in our previous paper; see \cite{Klapcsik2023}. The model is implemented in \textit{Python} and at each environment step it was solved by the initial value problem solver LSODA included \textit{SciPy} computational library. The absolute and relative tolerance was set to $10^{-10}$. To enhance computational speed, the \textit{numba} \textit{JIT} compiler \cite{numba-docs} was used to translate the ODE functions implemented in \textit{Python} to optimized machine code at runtime.

\subsection{Overview of the dynamical features}

\begin{figure} [!ht] 
	\centering
	\includegraphics[width=0.5\textwidth]{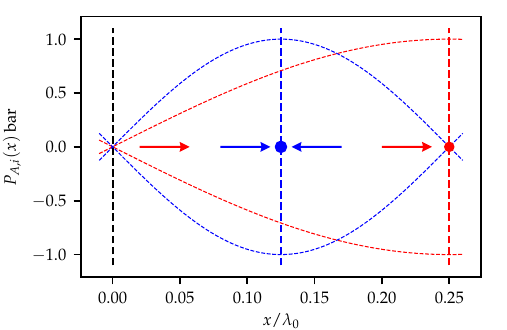}
	\caption{Schematic illustration of the acoustic field and the expected direction of the translational motion by linear theory.} 
	\label{Fig:LinearTheory}
\end{figure}

According to the linear theory, in a weak acoustic field, bubbles are attracted to either the pressure nodes or antinodes depending on their resonance size and excitation frequency \cite{Crum1975}. In the present study, the bubble size ($R_0=60\, \mathrm{\mu m}$) is chosen to be below the resonant sizes corresponding to the excitation frequencies ($R_{0,Res}=131\, \mathrm{\mu m}$ at $f_0=25\, \mathrm{kHz}$ and $R_{1,Res}=66\, \mathrm{\mu m}$ at $f_1=50\, \mathrm{kHz}$); thus, the bubble is attracted by the pressure antinodes, indicated by the arrows in Fig.~\ref{Fig:LinearTheory}. The pressure amplitude corresponding to the lower and higher frequencies are coloured by red and blue, respectively. The filled circles denote the antinodes. An antinode corresponding to the higher frequency component is located at $x/\lambda_0=0.125$. Below this threshold, between black and blue dashed lines, by increasing either $P_{A0}$ or $P_{A1}$, the expected translational direction is positive (blue and red arrow). The antinode corresponding to the lower frequency is located at $x/\lambda_0=0.25$. Between the blue and red dashed lines, the translational direction is either positive or negative depending on the magnitudes of the pressure amplitudes. 

To get an insight into the non-linear translational dynamics, one-dimensional bifurcation diagrams were calculated at fixed $P_{A1}$ values by changing $P_{A0}$ between $0.1\, \mathrm{bar}$ and $0.6\, \mathrm{bar}$ with a resolution of 151 steps. At every parameter combination, 5 initial positions were prescribed in the range of $x/\lambda_0\in(0.05, 0.2)$, and then initial value problem computations were carried out. The first 8192 acoustic cycles were treated as transients and these results were discarded. Only the (converged) trajectory segments obtained during the last 256 acoustic cycles were evaluated.

\begin{figure} [!ht] 
	\centering
	\includegraphics[width=0.5\textwidth]{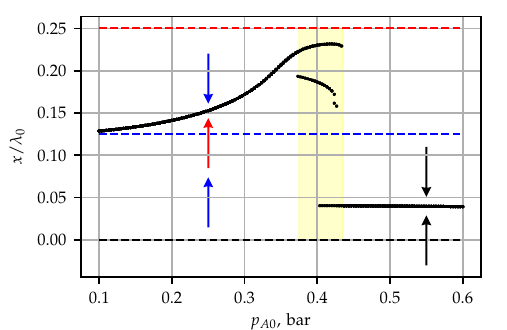}
	\caption{Mean bubble positions as a function of the pressure amplitude $P_{A0}$ at constant $P_{A1}=0.1\, \mathrm{bar}$. The horizontal dashed lines correspond to the pressure antinodes and nodes. The arrows indicate the direction of the translational motion.} 
	\label{Fig:SimpleBifurcation}
\end{figure}

An example is given in Fig.~\ref{Fig:SimpleBifurcation}, where the mean dimensionless displacement of the bubble is plotted as a function of the pressure amplitude $P_{A0}$. The horizontal dashed lines correspond to the pressure antinodes and nodes, see again Fig.~\ref{Fig:LinearTheory}. At low-pressure amplitude, the bubble translates to an intermediate equilibrium position between the antinodes. This is in agreement with the linear theory; thus, the expected directions can be figured out according to Fig.~\ref{Fig:LinearTheory}. The increasing pressure amplitude results in co-existing solutions (highlighted by the yellow rectangle). In addition, above pressure amplitude $P_{A0}=0.43\, \mathrm{bar}$, only one stable solution exists around $x/\lambda_0=0.4$. \textit{Thus, the combination of the dual frequency in the highly nonlinear regime may induce Bjerknes force that points in the opposite direction (black arrows) compared to the linear theory}. The further increase of the pressure amplitude causes oscillatory solutions \cite{Klapcsik2023}.
The above observation implies that \textit{the bubble position can be controlled arbitrarily between approximately $x/\lambda_0=0.05$ and $x/\lambda_0=0.25$ by the proper tuning of the pressure amplitude components}. This simple example demonstrates that exploiting non-linear dynamics allows more flexibility in position control. However, in the general case, where both $P_{A0}$ and $P_{A1}$ are tuned, the process is highly non-trivial. Thus, the aim is to develop a control system capable of arbitrarily positioning the bubble from any randomized initial position to any randomized final position. The idea is to use a deep neural network trained via reinforcement learning, to develop such a controller.

\section{The reinforcement learning framework}

To apply reinforcement learning, the above-described position control problem is formulated as a Markov decision process (MDP). The main elements of an MDP are the state, the action and the reward function \cite{sutton2018reinforcement, jeanfrancois2003markov}. $S$ and $A$ represent the state space and the action space, respectively and $R:S \times A\times S \rightarrow \mathbb{R}$ is the reward function. At each step, the agent observes the state $s\in S$ of the environment and chooses an action $a\in A$. One step is 50 acoustic cycles. Then, the environment moves to a new state $s'\in S$, and the agent receives the reward signal $r=R(s, a, s')$ from the environment. The reward function depends on the actual state of the environment, the action taken at that state, and the next state of the environment. The function that maps states to actions is the policy $a=\pi(s)$. The goal of the agent is to find a policy that maximizes the expected discounted return $R=\sum_{k=0}^{\infty}\gamma^k r_{k}$. The discount factor $\gamma \in (0,1)$ ensures the convergence of the infinite sum of reward. In the present paper, the agent controls the pressure amplitudes, and the environment is the bubble itself in the acoustic field. The elements of the MDP for the present problem are discussed below.

\textit{State} $S$ represents the observable quantities for the agent at each timestep. In the present paper, the state is assumed to be partially observable; thus, only bubble position values are observed. The first quantity is the desired (target) $x_T$ position in a given trial (episode). The next observed quantity is the actual position of the bubble $x_t$. To help the agent infer the direction and the velocity of the movement, the previous position value $x_{t-1}$ is also encoded in the actual observation. In the present paper, the maximum displacement of the bubble is $\lambda_0/4$; thus, the position values are normalized by the maximum of the observation limit to rescale the numerical values in the range of $[0, 1]$. Hereby, the state vector is defined as

\begin{equation}
	s = \left[\dfrac{4x_T}{\lambda_0}, \dfrac{4x_t}{\lambda_0}, \dfrac{4x_{t-1}}{\lambda_0} \right].
\end{equation}

\textit{Action Space} $A$ represents a set of actions that are available for the agent. According to the present control problem, the pressure amplitude values are changed in discrete timesteps; thus, the action space dimension is 2, where each value represents one of the two components of the pressure field

\begin{equation}
	a = [P_{A0}, P_{A1}].
\end{equation}

\noindent The pressure amplitude values are in the range of $P_{A0}, P_{A1}\in [P_{A,min}, P_{A,max}]$. Note that the pressure amplitude values are a piecewise constant function of time. One step of the environment represents 50 excitation periods ($50\tau$). Within each of these intervals, the pressure amplitude values are fixed. This time domain is higher than the transient regime required to establish a standing wave pattern in typical reactor scales; thus, the omission of complete acoustic simulation is a reasonable simplification.

The immediate \textit{reward} $r$ is the only feedback from the environment for the agent that represents how bad or good the last action was. To minimize the position error, the reward signal is defined as a shaped reward function as

\begin{equation}
	r_1 = 1 - \left(\dfrac{d}{d_{max}}\right)^k,
\end{equation}

\noindent where $d=|x_T - x_t|$ is the distance between the actual and the target position, $d_{max}$ is the maximum (possible) distance defined as 
\begin{equation}
	d_{max}=\max\left(x_{max}-x_T, x_T-x_{min}\right),
\end{equation}
and $k=0.2$ is the exponent, and the observation limits are $x_{min}=0.05\lambda_0$ and $x_{max}=0.25\lambda_0$. 

\begin{table*}[!ht]
	\centering
		\begin{tabular}{llr}
			Parameter &Notation& Value \\
			\hline
			Equilibrium bubble radius & $R_0$ & $60\, \mathrm{\mu m}$ \\
			Low frequency component & $f_0$ & $25\, \mathrm{kHz}$ \\
			High frequency component& $f_1$ & $50\, \mathrm{kHz}$ \\
			Observation space dimension& & $(3,1)$ \\
			Observation space limits& $[x_{min}, x_{max}]$ & $[0.05, 0.25]\cdot\lambda_0$ \\
			Action space dimension& & $(2,1)$ \\
			Action space limits& $[P_{A,min}, P_{A,max}]$ & $[0, 1\, \mathrm{bar}]$ \\
			Action duration &$\Delta\tau$ & $50\tau$ \\
			Episode length & $N_t$ & 12$\Delta\tau$ \\
            \hline
		\end{tabular}
    \caption{\label{tab:envparams} Properties of the environment.}
\end{table*}

The environment is implemented as an \textit{OpenAI GYM}-like environment \cite{openai_gym}, which is built on top of the model implementation presented in \textit{Section II}. The chosen algorithm used to train the agent is the deep deterministic policy gradient (DDPG) \cite{Silver2014, Lillicrap2016}. It is an off-policy, actor-critic algorithm that uses deep function approximations. DDPG can learn policies in continuous state and action spaces. The algorithm is implemented in Python by using the Pytorch deep-learning framework \cite{pytorch}. The implementation is verified by comparing the performance with the implementation of \textit{Clean RL} \cite{Gao2019Cleanrl} on benchmark problems provided by the \textit{Gym API}. Although, the above formalism allows customizable properties (frequency components, bubble size, time-step length, etc.) of the environment; the present paper demonstrates position control for one specific set of parameters, which are summarized in Table~\ref{tab:envparams}.

\subsection{Training of the agent}

At the beginning of the episodes, the bubble was assumed to be at rest, and the initial position was randomly prescribed ($R=R_0$, $\dot{R}=0$). Each episode terminates if the agent reaches the maximum allowed steps per episode or the target position is reached with a tolerance of $E_{x/\lambda_0}=0.01$. At every step, the experience ($e_t=(s_t, a_t, r_{t}, s_{t+1})$) is collected in the replay buffer. Batches of experiences are chosen randomly from the replay buffer to optimize the policy network according to the DDPG algorithm \cite{Lillicrap2016}. To ensure the exploration, a Gaussian noise is applied. The applied hyperparameters are summarized in Table~\ref{tab:hparams}. Keeping the hyperparameters fixed, the size of the neural network is optimized.

The policy network (actor) has as many input neurons as the size of the state vector. The output layer has two nodes for the two amplitude values with an activation function of tangent hyperbolic to scale the output values ($\hat{y}_i$) in the range of [-1,1]. Then, it is rescaled to actions (pressure amplitude) as 

\begin{equation}
	a_i = P_{Ai}= \dfrac{P_{A,max}-P_{A,min}}{2} \cdot \hat{y}_i + \dfrac{P_{A,max}+P_{A,min}}{2}.
\end{equation}

\noindent The critic network has $N+2$ input neurons (state and actions) and 1 output neuron for the action value.

\begin{figure} [!ht] 
	\centering
	\subfloat{\includegraphics[width=0.5\textwidth]{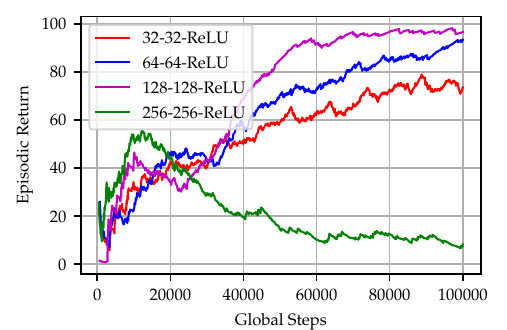}} \\
    \subfloat{\includegraphics[width=0.5\textwidth]{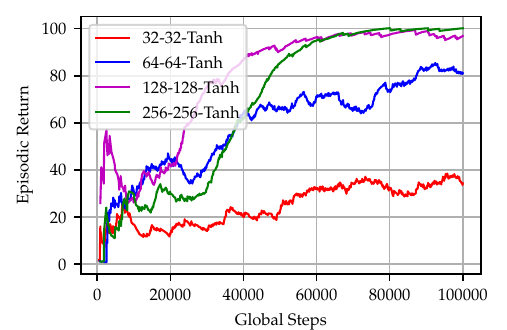}}
	\caption{The episodic return as a function of the (global) timesteps for ReLU and Tanh activation in the hidden layers. } 
	\label{Fig:Training}
\end{figure}

The optimal number of hidden layers and the number of neurons per layer are optimized by gradually increasing the network complexity and evaluating the training performance. Figure~\ref{Fig:Training} shows the smoothed (exponential moving averaged) of the episodic return as a function of training steps for different network architectures and activation functions. Note that the theoretical maximum for an infinite time horizon is $R_{max}\approx\lim_{r\rightarrow\infty}{\sum_{k=0}^{\infty}\gamma^k r_{k}}=100$. The top panel shows the episodic reward as a function of steps for different network structures with a rectified linear unit (ReLU) as activation in the hidden layers. The highest reward and the fastest convergence are achieved using two hidden layers with 128 neurons per hidden layer. The increasing model complexity resulted in poor performance. It is worth mentioning that with one hidden layer, the algorithm is not converged (not shown here) for the present problem. Using tangent hyperbolic (Tanh) as activation in the hidden layers, slightly better convergence is achieved (see the bottom diagram).

\begin{table} [!hb]
    \centering
	\begin{tabular}{lr}
		Hyperparameter & Value \\
		\hline
		Learning rate & $2.5\cdot 10^{-4}$ \\
		Discount factor & $0.99$ \\
		Batch size & $256$ \\
		Buffer size & $10\, 000$ \\
		Soft update weight ($\tau$) & $5\cdot 10^{-3}$ \\
		Exploration noise & 0.1 \\
        \hline
	\end{tabular}
    \caption{\label{tab:hparams} Hyperparameters of the agent. }
\end{table}

\begin{figure} [!ht]
    \centering
    \includegraphics[width=0.5\textwidth]{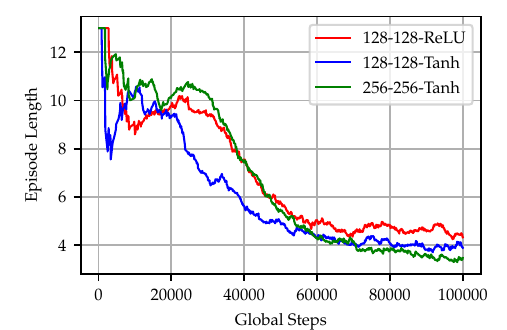}
    \caption{The episode length as a function of the (global) timesteps for the three best models.}
    \label{Fig:ControlTime}
\end{figure}

The maximization of an improperly specified reward may lead to false policy from the point of view of the main objective, which is to achieve position control within the shortest time. Therefore, the episode length (exponential moving averaged) for the top three models is plotted in Figure~\ref{Fig:ControlTime}. The results show that shorter episodes, i.e., fastest position control can be achieved using Tanh activation in the hidden layers.

\section{Case studies}

\begin{figure} [!ht]
    \centering
    \includegraphics[width=0.5\textwidth]{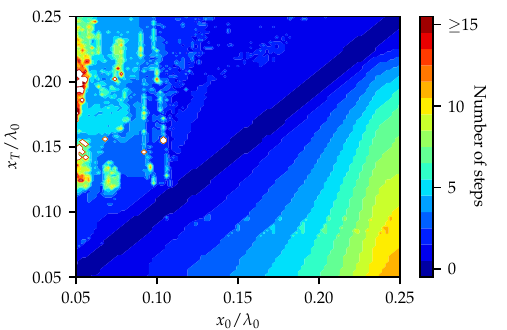}
    \caption{The number of steps required to drive the bubble from initial position $x_0/\lambda_0$ to target position $x_T/\lambda_0$.}
    \label{Fig:ModelPerformance}
\end{figure}

For further investigations, the trained model with the highest complexity (256-256-Tanh) is chosen. To evaluate the model performance, numerous simulations were carried out with various initial and target positions. The pressure amplitudes were chosen according to the trained policy. The results are visualized in Figure~\ref{Fig:ModelPerformance}, where the total number of steps required for position control is plotted as a function of the initial $x_0/\lambda_0$ and target $x_T/\lambda_0$ bubble positions. The applied resolution is $x_0\times x_T = 101\times101$. The figure shows that the position control was successful in the majority of the parameter space (97\%), and the bubble was driven to the target position in less than 15 steps, i.e.; less than $15\times50=750$ acoustic cycles. 

\begin{figure} [!ht]
    \centering
    \subfloat{\includegraphics[width=0.5\textwidth]{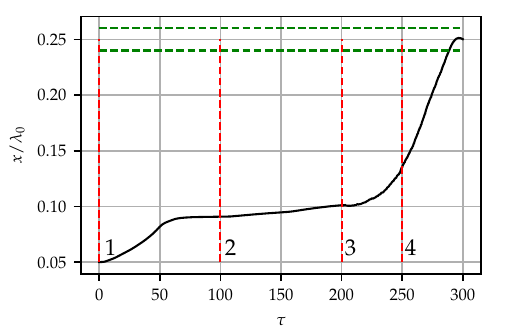}} \\
    \subfloat{\includegraphics[width=0.5\textwidth]{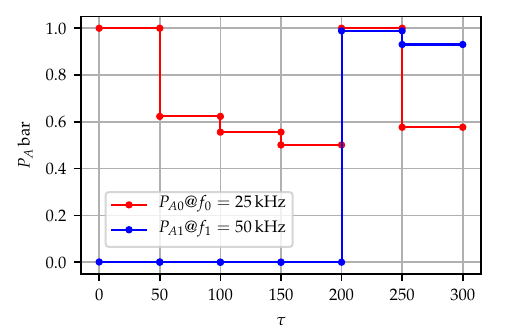}}
    \caption{The trajectory of the bubble controlled by the RL agent (top panel) and the chosen pressure amplitude values (bottom panel).}
    \label{Fig:Trajectory}
\end{figure}

Figure~\ref{Fig:Trajectory} shows a specific test case when the bubble is driven from one side of the space domain to the other. On the top panel, the trajectory of the bubble (black curve) is plotted as a function of the acoustic cycles, while the bottom panel shows the applied pressure amplitude values. The green horizontal dashed lines denote the thresholds for successful control around the target position $x_T/\lambda_0 = 0.25$. The trajectory shows 3 different kinds of segments. At $x/\lambda_0=0.05$ the agent chooses high $P_{A0}$ and zero $P_{A1}$ pressure amplitude. The resulting pressure field pushes the bubble rapidly around $x/\lambda_0=0.082$. Between $x/\lambda_0=0.082$ and $x/\lambda_0=0.1$, the agent chooses moderate and decreasing $P_{A0}$ while $P_{A1}$ is kept at zero. These actions are justified according to the linear Bjerkness theory \cite{Mettin1997, Matula1997, Akhatov1997, Louisnard2008}. The bubble moves to the pressure node located at $x/\lambda_0 = 0.25$. However, there is a limit for pressure amplitude $P_{A0}$. Our preliminary parameter study (see Fig. 2 panel A in paper \cite{Klapcsik2023}) revealed that an intermediate equilibrium solution exists at bubble size $R_0=60\, \mathrm{\mu m}$. Therefore, the pressure amplitude must be as high as possible to ensure short control time but with care for the attraction of the intermediate equilibrium. Above $x/\lambda_0 = 0.1$, the agent chooses mixed pressure amplitude combinations that result in fast and precise position control.

\begin{figure*}[!ht]
    \centering
    \subfloat{\includegraphics[width=0.5\textwidth]{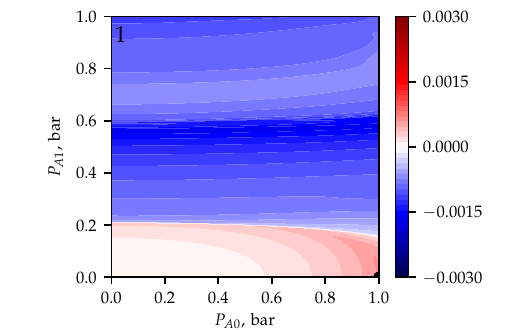}}
    \subfloat{\includegraphics[width=0.5\textwidth]{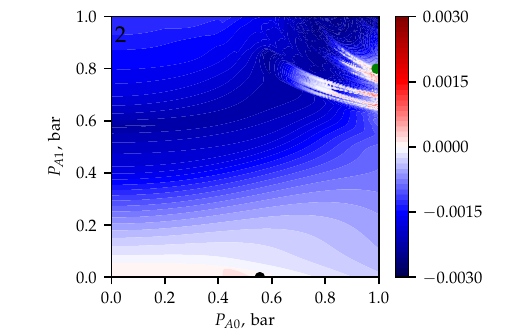}} \\
    \subfloat{\includegraphics[width=0.5\textwidth]{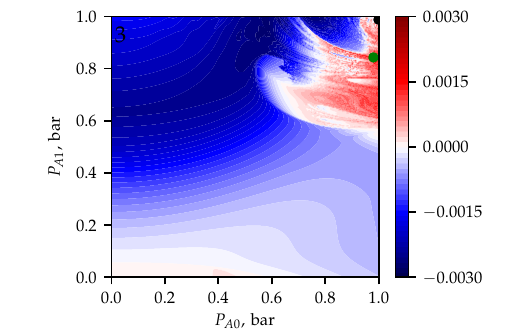}}
    \subfloat{\includegraphics[width=0.5\textwidth]{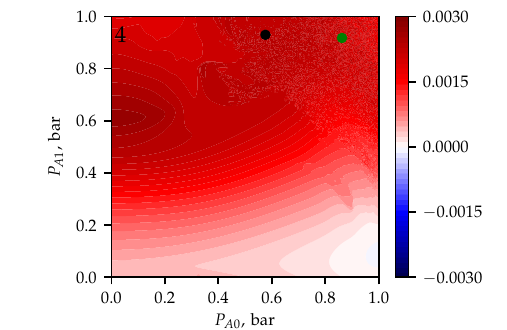}}
    \caption{The averaged translational velocity over 50 acoustic periods as a function of the pressure amplitude components. The black dots denote the optimal parameter set obtained by the direct search for the maximum velocity and the green dots denote the parameter set chosen by the RL agent.}
    \label{Fig:Actions_BruteForce}
\end{figure*}

To verify the actions of the agent, 2-dimensional parameter studies were carried out in the parameter space of the pressure amplitudes via GPU-accelerated initial value problem computations to directly search the optimal pressure amplitude combinations. The applied solver was written in \textit{Python} and using the \textit{numba} library that supports writing CUDA kernels \cite{numba-docs}. The implemented algorithm is the fourth-order Runga-Kutta-Cash-Karp method with fifth-order error estimation. At every pressure amplitude combination, numerical simulations were carried out over 50 acoustic cycles (one time-step) and the averaged translational velocity was calculated. During the computations, the absolute and relative tolerance were set to $10^{-10}$. These parameter maps were calculated along the trajectory given in the top panel of Fig.~\ref{Fig:Trajectory} at bubble positions labelled from 1 to 4 and marked with vertical red dashed lines. The results are given in Figure~\ref{Fig:Actions_BruteForce}. The resolution of each parameter map is $P_{A0}\times P_{A1} = 256\times256$. The black dots denote the maximum velocity. The parameters chosen by the agent are plotted with green dots.

In the first step, the direct search and agent suggest the same action that results in maximal displacement. In the second case, the direct search suggests an optimal parameter set at ($P_{A0}=1\, \mathrm{bar}$ and $P_{A1}=0.8\, \mathrm{bar}$). Although it results in a higher displacement compared to the trajectory obtained by the RL, this narrow domain requires an overly precise amplitude setting that may be infeasible in practice. \textit{The RL chose a more robust decision compared to the direct search}. As the bubble moves towards the positive direction, the domain resulting in a high translational velocity increases. Thus; in the third case, the agent picks a parameter set from this domain. Lastly, the agent picks an action that results in an accurate final position. 

\section{Conclusion}

The present paper proposes that reinforcement learning (RL) can be a powerful tool in developing controllers to manipulate bubble clusters (e.g., minimising attenuation via manipulating bubble positions to keep an optimal structure). To demonstrate this a simple but non-trivial case is investigated, where the task of the agent was to control a single bubble on the shortest path from any random position to the target position. In our study, the target position was randomized. However, depending on the applications, the target position can be specified, e.g., the best spot of chemical activity \cite{Islam2019b, Rashwan2019, Gedanken2004}, or in the case of ultrasonic water treatment \cite{Jose2010, Esclapez2010, Zupanc2013, Dular2016} or clot lysis \cite{Bader2015, Acconcia2013} the bubble may be driven near specific surfaces. It is worthwhile to mention that the agent was rewarded only for the minimization of the distance, which implicitly encouraged the agent to seek fast position control. A fine-tuned reward function allows encoding additional constraints such as minimization of the consumed energy \cite{Sojahrood2020b, Sojahrood2020c, Sojahrood2021, Sojahrood2021b} or avoiding bubble break-off due to surface instabilities \cite{Brenner1995, Bogoyavlenskiy2000, Dollet2008, Hao1999, Holzfuss2008, Lalanne2015, Shaw2006, Shaw2009, Shaw2017} via adding penalties or the maximization of various chemical products \cite{Islam2021, Pflieger2015, Ouerhani2015, Xu2013, Islam2019a, Merouani2015, Kerboua2019}.

The RL agent was capable of finding optimal control on a wide range of initial positions that move the bubble to an arbitrary target position within the shortest amount of time. For example, in the case presented in Fig.~\ref{Fig:Trajectory}, the naive approach (pushing the bubble with the primary Bjerknes force) would require 2150 acoustic periods, which is 7.2 times higher than the solution of the RL. It was capable of developing non-trivial solutions, see Fig.~\ref{Fig:ModelPerformance}. In the lower right domain, the bubble is moved from the stable antinode close to the unstable node ($x/\lambda=0.05$). That direction of movement can not be justified by the linear theory of translational bubble motion (see again Fig.~\ref{Fig:SimpleBifurcation}).

The structural similarities of parameter maps observed in our previous paper \cite{Klapcsik2023} imply that the size of the bubble $R_0$ determines the proper choice of the $f_0$, and $f_1$ frequency components. To develop universal control one can extend the observation vector with an equilibrium bubble size and apply for example parameterized action space \cite{Masson2015, Delalleau2019, Fan2019}, such as $a= [(f_0, f_1)_k, (P_{A0}, P_{A1})_k]$, where $(f_0, f_1)_k\in[(f_0, f_1)_1, (f_0, f_1)_2 ... (f_0, f_1)_K]$ is a discrete frequency combination (assuming standing waves) and $(P_{A0}, P_{A1})_k$ is its parameter vector. The investigation of such cases was beyond the scope of the present paper.

The high-resolution simulations revealed that by precise parameter tuning, better trajectories can be found via direct numeric solutions. However, implementing real-time controllers, for more complex problems, may be infeasible due to the increasing computational time. The computational time of a single parameter map plotted in Fig.~\ref{Fig:Actions_BruteForce} is approximately 4-5 seconds on a GeForce GTX 1070 Graphics card. The applied GPU-accelerated solver has a similar computation performance as the MPGOS \cite{MPGOS_GitHub, Hegedus2019a} program package (not shown here). On the contrary, the inference time of the neural network (policy) is approximately $1.2\, \mathrm{ms}$ on the same GPU (including the data transfer between host and device). The computation time of the direct search method scales unfavourably by the increasing complexity, e.g., adding a secondary bubble compared to the network inference.

The increasing complexity of the problem (e.g., multi-frequency driving \cite{Suo2018}, extension to bubble cluster \cite{Doinikov2004b, Mettin2007, Mettin2005, Mettin1999b, Mettin1999, Koch2003}) requires more sophisticated reinforcement learning algorithms, that allow the sufficient exploration of the state and action space. One promising candidate is Proximal Policy Optimization (PPO) \cite{Schulman2017}, known for its scalability using vectorized environments allowing efficient parallelization \cite{heess2017emergence, rudin2022learning} and improved sample efficiency in training.





\section*{Declaration of Competing Interest}
\noindent The authors declare that they have no known competing financial interests or personal relationships that could have appeared to influence the work reported in this paper.

\section*{Acknowledgements}
The research reported in this paper is part of project No. BME-NVA-02, implemented with the support provided by the Ministry of Innovation and Technology of Hungary from the National Research, Development and Innovation Fund, financed under the TKP2021 funding scheme. 
The research was supported by the J\'anos Bolyai Research Scholarship (BO/00217/20/6) of the Hungarian Academy of Sciences and by the New National Excellence Program of the Ministry for Culture and Innovation from the source of the National Research (\'UNKP-22–5-BME-310) and by NVIDIA Corporation via the Academic Hardware Grants Program and by the European Union project RRF-2.3.1-21-2022-00004 within the framework of the Artificial Intelligence National Laboratory.
The authors acknowledge the financial support of the Hungarian National Research, Development and Innovation Office via NKFIH Grants OTKA PD 142254 and OTKA FK 142376. This project has received funding from the European Union's Horizon research and innovation programme under the Marie Skłodowska-Curie grant agreement No.\,101064097.

\bibliography{pre_kk_hf_gytb_j}

\end{document}